\documentclass[%
 aip,
 amsmath,amssymb,
preprint,%
]{revtex4-1}

\usepackage{graphicx}
\usepackage{dcolumn}
\usepackage{bm}
\usepackage[mathlines]{lineno}

\usepackage[utf8]{inputenc}
\usepackage[T1]{fontenc}
\usepackage{mathptmx}
\usepackage{etoolbox}

\usepackage{amsmath}
\usepackage{color}
\usepackage[caption=false]{subfig}
\usepackage{siunitx}
\usepackage[version=4]{mhchem}

\makeatletter
\def\@email#1#2{%
 \endgroup
 \patchcmd{\titleblock@produce}
  {\frontmatter@RRAPformat}
  {\frontmatter@RRAPformat{\produce@RRAP{*#1\href{mailto:#2}{#2}}}\frontmatter@RRAPformat}
  {}{}
}%
\makeatother
\begin{document}

\preprint{AIP/123-QED}

\title[Effects of non-condensable gas on laser-induced cavitation bubbles]{Effects of non-condensable gas on laser-induced cavitation bubbles}

\author{D. B. Preso}
\affiliation{Institute of Mechanical Engineering, École Polytechnique Fédérale de Lausanne, Avenue de Cour 33 Bis, 1007 Lausanne, Switzerland}
\email{davide.preso@epfl.ch}

\author{D. Fuster}%
\affiliation{Sorbonne Université, Centre National de la Recherche Scientifique, UMR 7190, Institut Jean Le Rond $\partial$’Alembert, F-75005 Paris, France}

\author{A. B. Sieber}
\affiliation{Institute of Mechanical Engineering, École Polytechnique Fédérale de Lausanne, Avenue de Cour 33 Bis, 1007 Lausanne, Switzerland}

\author{M. Farhat}
\affiliation{Institute of Mechanical Engineering, École Polytechnique Fédérale de Lausanne, Avenue de Cour 33 Bis, 1007 Lausanne, Switzerland}

\date{\today}

\begin{abstract}
This study presents experimental observations of single laser-induced cavitation bubbles collapsing in water with different levels of air saturation.
The average trends of the bubble size reveal a clear yet little dependence of the energy dissipation at collapse on the water gas content.
Similarly, the observed trend of luminescence energy at collapse varies within the investigated range of air saturation levels, in agreement with findings in similar works.
We argue that perturbations in bubble shape may amplify with decreasing air saturation, consequently affecting light emission.
\end{abstract}

\maketitle

The presence of dissolved gas in flowing liquids lowers their tensile strength and significantly influences cavitation occurrence \cite{BrennenBook,BlanderKatz,Plesset,Amini}. 
However, our understanding of the effect of the gas on cavitation bubble dynamics, the gas composition inside the bubbles, and the behaviour of the gas at the bubble-liquid interface, remains limited.
Moreover, cavitation bubbles likely contain a mixture of condensable and non-condensable components, increasing the complexity for their investigation.
To date, owing to the small scale both in time and space, direct probing of the bubble contents remains unrealistic, making only indirect measurements possible \cite{BrennerHilgenfeldtLohse,FlanniganHopkinsSuslick,AkhatovLindauLauterborn,KellerMiksis}.
A comprehensive understanding of the effect of the gaseous content on cavitation bubble collapse is essential for the development of innovative cavitation-based technologies \cite{SuslickMdleleniRies}, as well as for the improvement of advanced numerical models.
\par The role of dissolved gases on cavitation bubbles sparked considerable interest in the nineties, particularly in relation to studies on single bubble sonoluminescence.
Several researches highlighted the importance of the nature and composition of the gaseous phase on the light emission from periodically-driven gas bubbles trapped in an acoustic field \cite{BarberWuPutterman,GaitanCrumRoy,Hiller,Brujan,Gompf,Krefting,Young,Toegel,FlanniganSuslick}.
However, the investigation of these bubbles is constrained to a limited parameter space, involving a narrow range of dissolved gas concentrations \cite{BrennerHilgenfeldtLohse,Wolfrum}.
Moreover, these bubbles allow long time scale phenomena, such as mass diffusion of dissolved gases, to become dominant \cite{LohseBrenner,Baghdassarian}.
Here, the study of single transient cavitation bubbles, besides overcoming these challenges, is better suited for comparison with bubbles in cavitating flows \cite{Wolfrum,OhlLuminescence}.
\par Several numerical models have been proposed to elucidate the composition and behaviour of the gaseous phase within cavitation bubbles.
\citeauthor{Fujikawa} \cite{Fujikawa}, who modeled the dynamics of single laser-induced bubbles, evidenced the effect of non-equilibrium condensation on the bubble dynamics due to the interaction of condensable vapour and non-condensable gas within the bubble.
A similar model was also proposed by \citeauthor{AkhatovLindauLauterborn} \cite{AkhatovLindauLauterborn}, who reported that a small amount of non-condensable gas could significantly impact the bubble dynamics by hindering the condensation of vapours at the bubble-liquid interface.
Similar numerical models have also been developed more recently \cite{MagalettiMarinoCasciola,FusterMontel,Hao,Szeri}.
Differently, experimental investigation performed by \citeauthor{Baghdassarian} \cite{Baghdassarian} revealed no significant effects of the gas nature and concentration on the light emission at the collapse of single laser-induced bubbles.
Successively, \citeauthor{Wolfrum} \cite{Wolfrum} came also to the same conclusion.
However, despite these findings, this facet has been largely overlooked in experimental studies \cite{SupponenLuminescence}, and knowledge in this area remains limited \cite{Liu}.
\par This Letter aims to contribute to the ongoing discussion regarding the influence of non-condensable gases on the collapse of transient cavitation bubbles.
We present experimental results of the dynamics of single laser-generated cavitation bubbles in water, in which we systematically varied the saturation level of dissolved air.
Additionally, we collect and analyze the light emitted at the collapse of the bubbles.
The main message of this Letter is that, despite observing some minor effects, the dissolved gas at the concentrations investigated exerts a virtually negligible effect on the bubble dynamics.
\begin{figure}
    \centering
    \includegraphics[width=0.7\textwidth]{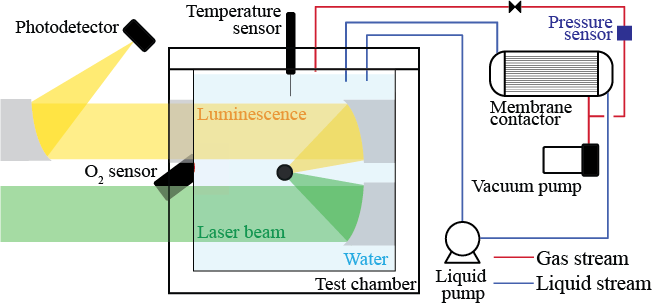}
    \caption{Schematic representation of the experimental setup. The oxygen meter is off the light beams' path.}
    \label{fig:setup}
\end{figure}
\par Our experimental setup, schematically illustrated in Figure \ref{fig:setup}, generates highly-spherical laser-induced cavitation bubbles using an immersed parabolic mirror \cite{ObreschkowQuest,SupponenDetailed,SieberSand,SieberGel}.
The maximum bubble radius $R_0$ is approximately $1.7$ mm, with a variation of $\pm 4$\% due to natural laser energy oscillation.
Such bubbles are weakly influenced by the hydrostatic pressure gradient and nearby solid surfaces, with an anisotropy parameter value of $\zeta \approx 4 \times 10^{-4}$, and thus their collapse is largely spherical.
Details about $\zeta$ can be found in \citeauthor{ObreschkowUniversal} \cite{ObreschkowUniversal} and \citeauthor{SupponenScaling} \cite{SupponenScaling}.
The bubble is generated at the center of a transparent cubic test chamber, whose volume is large enough not to affect the bubble dynamics.
The gas-tight test chamber is filled with distilled water and connected to a degassing system that controls the air saturation level in the water in a closed-loop configuration.
The system includes a pump that recirculates water through the lumenside of a membrane contactor (3M Liqui-Cel Series), permeable by gases via a pressure-driven process.
The pressure in the shellside $p^*$ of the contactor is adjusted with a vacuum pump and controlled with a pressure sensor.
At equilibrium, the saturation level of water, expressed in molar concentration $M^*$ (moles of air per unit volume of water), is proportional with $p^*$ according to Henry's law.
The saturation level of water is monitored with a temperature-compensated oxygen meter (PreSens Fibox 3), whose sensor is placed in the test chamber at the same height as the bubble spot.
During degassing operations, the pressure of the test chamber is reduced to limit diffusion of gas back into the water.
Ambient pressure is only restored during the experimental timeframe, which remains short enough to prevent back-diffusion of air from the free surface down to the bubble location.
Owing to the similar permeability of \ce{O2} and \ce{N2} through the membrane, we assume that the degassing rate of oxygen in the water is similar to the one of air.
In this case, we can monitor the air saturation level by measuring the oxygen concentration in the water.
In any case, after reaching oxygen equilibrium concentration, we extend the degassing process to ensure equilibrium for \ce{N2} as well.
The temperature $T_\infty$ of the water is maintained relatively constant, fluctuating between 24.6 and \SI{24.8}{\celsius}.
\par We resolve the bubble dynamics from shadowgrams captured at a rate of $5 \times 10^5$ frames s$^{-1}$ and an exposure time of $200$ ns with a high-speed camera (Shimadzu HPV-X2).
We define the potential energy of the bubble $E_{\text{p0}}$ as
\begin{equation}
    E_{\text{p0}} = (4/3) \pi R_0^3 \left(p_\infty - p_\text{v}\right),
\end{equation}
where $p_\infty$ is the liquid pressure at rest, which equals the ambient pressure $p_\infty$ of 96.6 kPa, and $p_\text{v}$ the water vapour partial within the bubble, assumed equivalent to the vapour pressure of the water $p^*_\text{v}(T_\infty)$.
The latter is obtained from tabulated values \cite{Perry}.
Likewise, we define the potential energy of the first rebound bubble as $E_{\text{p1}} = (4/3) \pi R_1^3 \left(p_\infty - p_\text{v}\right)$, where $R_1$ is the maximum radius of the rebound bubble.
\par We also capture luminescence signals at bubble generation and  collapse using a photodetector (ThorLabs, DET10A/M Silicon Detector, 200-1100 nm wavelength, 1 ns rise time).
The light emitted is collected by a second immersed parabolic mirror, and transmitted through a fused silica window onto a third parabolic mirror outside the test chamber, which focuses the light onto the photodetector.
The silica window enhances ultraviolet transparency. 
This procedure is detailed elsewhere \cite{SupponenLuminescence}.
Owing to the small size of the bubbles and the ensuing value of $\zeta$, we neglect the upward bubble displacement effect, omitting it for photodetector signal correction \citeauthor{SupponenScaling}.
The light emitted at bubble generation and collapse provides an accurate measure of the bubble's lifetime.
Additionally, we compute the energy of the luminescence signal at collapse $E_\text{lum}$ as
\begin{equation}
    E_\text{lum} =  \frac{\psi}{Z} \int U_\text{ph}(t)^2 dt,
\end{equation}
where $U_\text{ph}(t)$ is the electric signal obtained from the photodetector, and $Z$ is a calibration constant which is not available for our experimental setup.
Consequently, $E_\text{lum}$ is reported in arbitrary units.
Assuming a uniform light emission in the solid angle 4$\pi$ sr, $E_\text{lum}$ is multiplied by a constant $\psi = 0.067^{-1}$ as the parabolic mirror collects only around 6.7\% of the total emitted light.
\par We utilize the Keller-Miksis model \cite{KellerMiksis} to fit the experimental data of the bubble oscillation up to the first rebound and estimate the bubble's internal pressure $p_\text{b}$.
The latter is described as $p_\text{b}(t) = p_\text{v} + p_\text{g0}\left(R_0/R\right)^{3\gamma}$, where $R(t)$ is the bubble radius, $p_\text{g0}$ is the non-condensable gas partial pressure within the bubble at maximum expansion, and $\gamma$ its heat capacity ratio.
A detailed explanation on fitting the bubble pressure from experimental data with the Keller-Miksis model is available elsewhere \cite{PresoRapids}.
\begin{figure}
    \centering
    \includegraphics[width=0.8\textwidth]{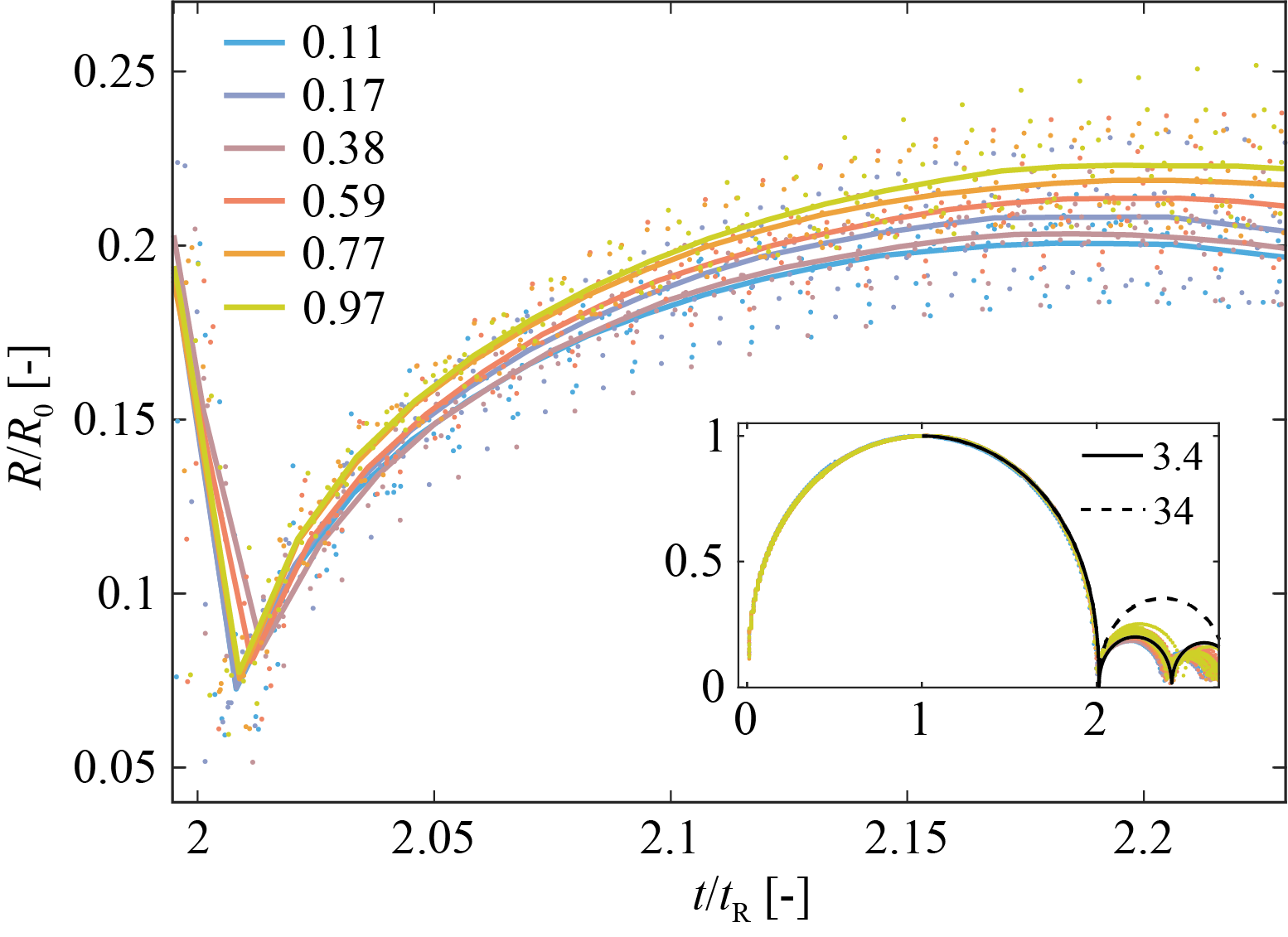}
    \caption{Radial evolution over time in normalized coordinates of single laser-induced cavitation bubbles generated in water with different air saturation levels $M^*/M_\infty$. 
    The inset plot shows the overall collapse and rebound process, whereas the main plot focuses on the final stage of the collapse and first rebound.
    The points are discrete data of 10 different experimental measurements.
    The solid lines display the average of the measurements. The color code indicates $M^*/M_\infty$.
    In the inset plot, the solid and dashed black lines refer to the solution of the Keller-Miksis models with two different $p_\text{g0}$ (in Pa), respectively.}
    \label{fig:rt}
\end{figure}
\par
\begin{figure}
    \centering
    \includegraphics[width=0.78\textwidth]{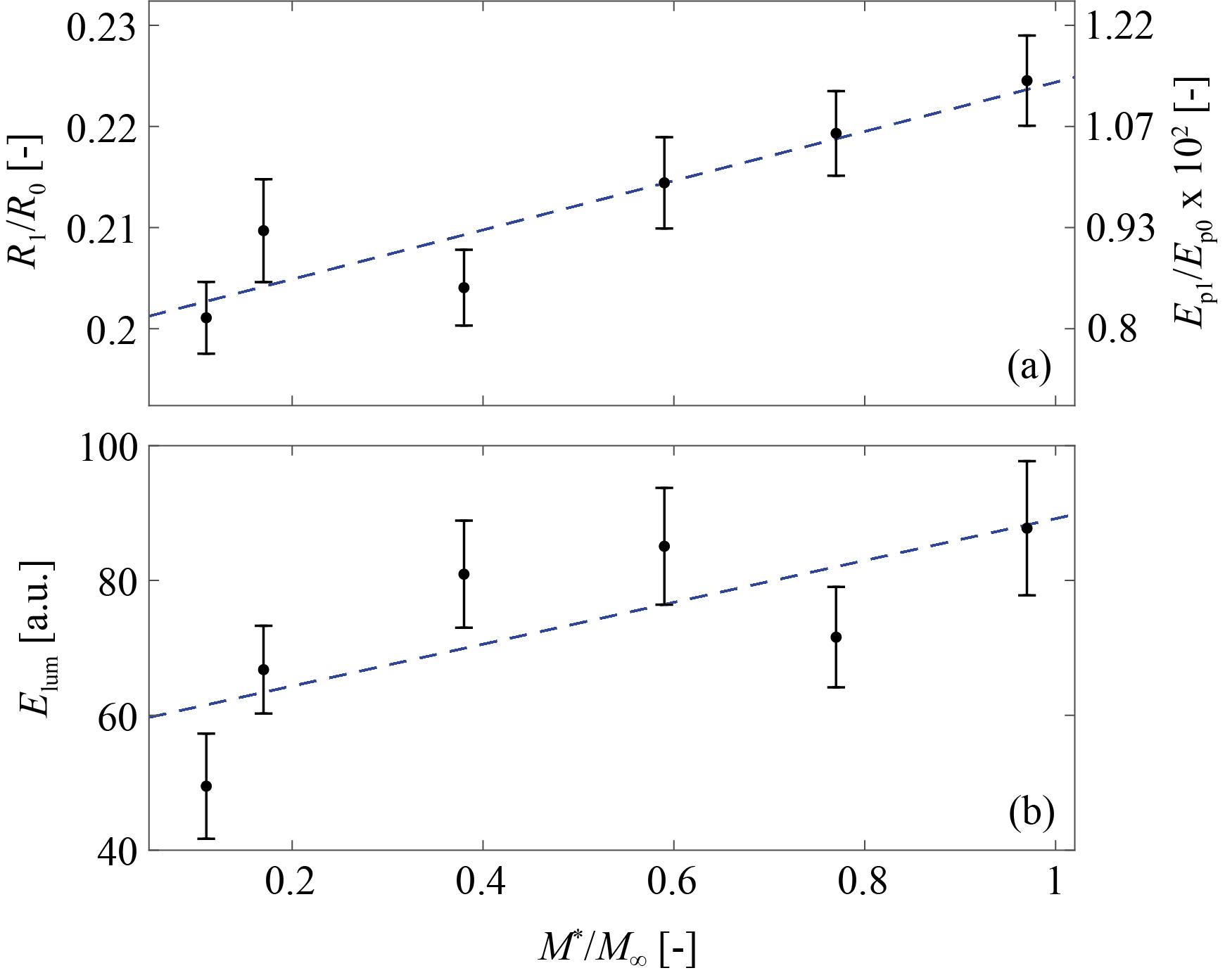}
    \caption{(a) Maximum radius (and corresponding energy on the right y-axis) of the rebound bubble, in normalized coordinates, and (b) luminescence energy at bubble collapse, as a function of the air saturation level of the water. For luminescence, the dots display the average value of 30 measurements, whereas the error bars the standard error of the measurements. In both (a) and (b), the blue dashed lines are linear regressions of the data for visualization purposes.}
    \label{fig:rebound}
\end{figure}

 Figure \ref{fig:rt} shows, in normalized coordinates, the temporal evolution of the radius of cavitation bubbles generated in water at different air saturation $M^*/M_\infty$, where $M_\infty$ is the saturation at equilibrium at $p_\infty$ and $T_\infty$.
Here, $t_\text{R}=0.915 R_0 \sqrt{\rho/p_\infty}$ is the Rayleigh collapse time for a single empty cavitation bubble \cite{Rayleigh}, with $\rho$ representing the water density.
The inset plot in Figure \ref{fig:rt} provides an overview of the bubble growth, collapse, and rebounds, while the main plot zooms in on the final stage of the collapse and the first rebound.
As displayed, the collapse appears nearly identical among all bubbles (10 measurements for each saturation level), although data of rebound bubbles are very disperse.
Nonetheless, the averaged radial evolution, depicted by the solid lines, reveals small differences of the rebound size depending on the air saturation level of the water.
From $M^*/M_\infty=0.11$ to $0.97$, the average maximum radius of the rebound bubble progressively increases by approximately 12\%.
This trend is more clearly represented in Figure \ref{fig:rebound}(a), which illustrates the normalized maximum radius of the rebound bubble as a function of $M^*/M_\infty$.
In terms of energy, the variation in rebound size is insignificant.
On average, bubbles generated in water at $M^*/M_\infty=0.97$ received only 0.3\% more $E_\text{p0}$ than those at $M^*/M_\infty=0.11$.
\par Regarding the collapse time, we observe no significant impact of the air saturation of the water, despite the accuracy of the photodetector.
\par Fitting the bubble's internal pressure from the experimental results with the Keller-Miksis model reveals that $p_\text{g0}$ varies on average between 3.4 and 5.4 Pa from the least to the most saturated water investigated.
Under the model's assumption, these results demonstrate that a tenfold augmentation in dissolved gas in the water does not translate into the same increase for $p_\text{g0}$.
This is also evident in the inset plot in Figure \ref{fig:rt}, where we plotted the Keller-Miksis model predictions with $p_\text{g0}=3.4$ Pa alongside the solution at 34 Pa.
\begin{figure}
    \centering
    \includegraphics[width=0.7\textwidth]{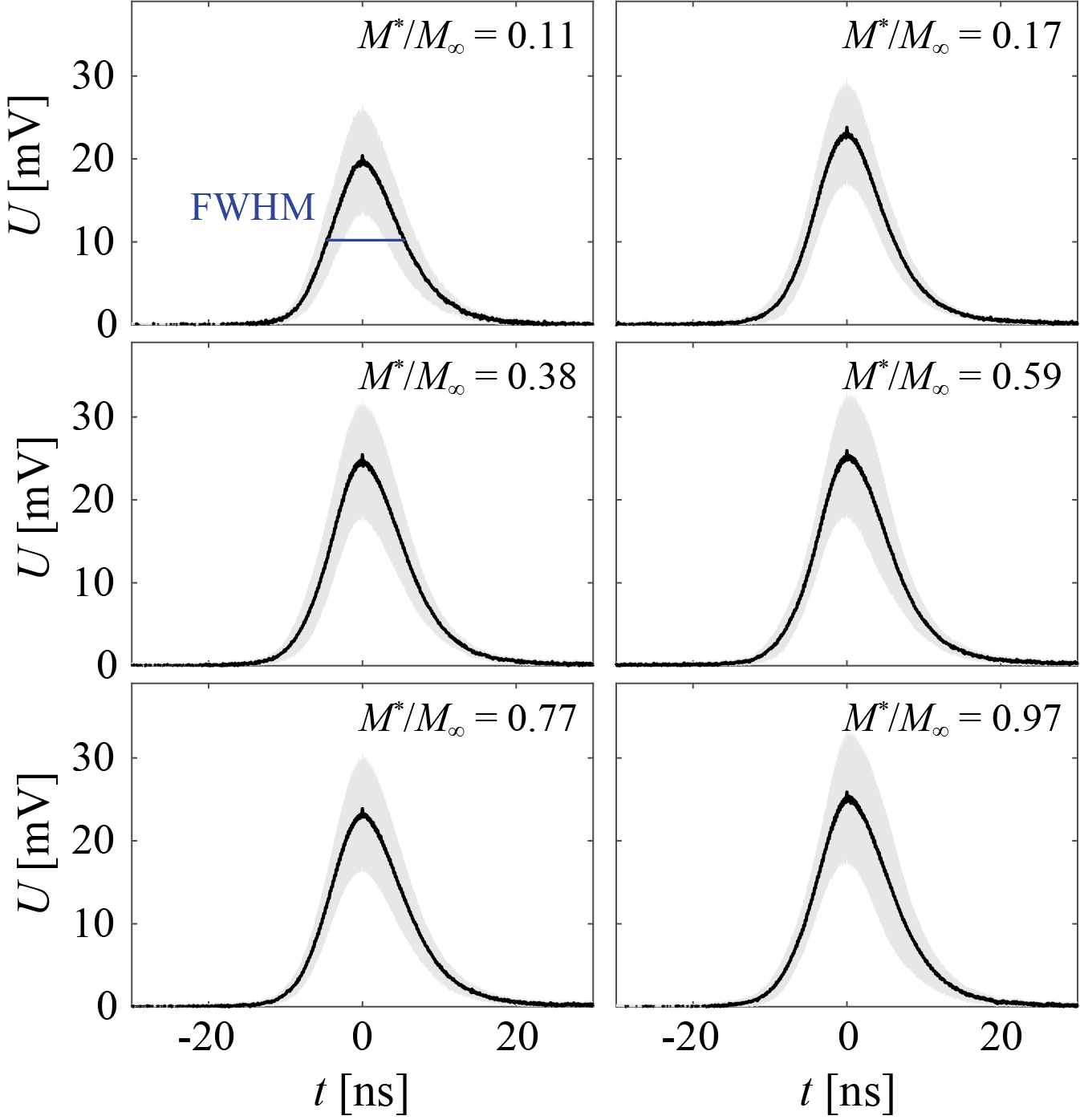}
    \caption{Luminescence signals for different air saturation levels $M^*/M_\infty$ of the water. The solid lines indicate the mean of 30 measurements, where signals are averaged by setting the peak at $t=0$. The shaded area shows the standard deviation of the measurements. The blue line in the top left plot shows the full-width at half maximum (FWHM) of the average signal.}
    \label{fig:signals}
\end{figure}
\par Although luminescence data are also considerably dispersed, the average luminescence signal energy also increases with rising air saturation level.
The trend is illustrated in Figure \ref{fig:rebound}(b), where, on average, $E_\text{lum}$ strongly increases from the lowest to the highest $M^*/M_\infty$ explored.
\par Raw luminescence signals collected by the photodetector are presented in Figure \ref{fig:signals}.
The averages of the photodetector signals at different air saturation levels are consistent with signals from similar bubbles reported in previous works \cite{SupponenLuminescence,Baghdassarian}, with a full width at half-maximum of approximately 10 ns.
A noteworthy observation is the detection of a single peak for every bubble, indicating a highly spherical collapse and light radiation from a single emission region \cite{OhlLuminescence,SupponenLuminescence,Sukovich}.
Moreover, it is interesting to observe the neat bell-like shape of the signal, also synonymous with spherical collapse.
\citeauthor{SupponenLuminescence} \cite{SupponenLuminescence} observed multiple peaks in the luminescence signal for deformed bubbles.
\par The significant scatter observed in rebound size and light emission at collapse (Figure \ref{fig:rt} and \ref{fig:rebound}) is likely due to the poorly repeatable phenomena occurring in the last stage of the collapse.
This phenomena are probably influenced by variable absorbed energy at bubble generation, initial plasma anisotropies, bubble asymmetries, liquid impurities, and bubble shape perturbations.
However, the average maximum rebound size and luminescence energy were found to be dependent on $M^*/M_\infty$.
The observed trends are most likely due to the variation of the partial pressure of non-condensable gas within the bubble, resulting from (i) vaporization of dissolved gas at initial plasma generation, (ii) plasma recombination, and (iii) gas diffusion from the liquid \cite{SupponenLuminescence,Liang}.
Gas generated from plasma recombination is proportional to the laser energy deposited at bubble generation \cite{SupponenRebounds,Sato} and we assume no variation with the gas saturation of the water. 
In a study by \citeauthor{Sato} \cite{Sato}, who investigated bubbles similar in size to those reported in this work, hydrogen resulting from plasma recombination was observed to be in the order of $10^{-11}$ grams.
The initiation plasma, on the other hand, vaporizes approximately $10^{-11}$ to $10^{-10}$ grams of air in water at $M^*/M_\infty$ ranging from around 0.1 to 1 using the approach of \citeauthor{Baghdassarian} \cite{Baghdassarian}.
Estimation of gas due to diffusion from the liquid into the bubbles suggests a total mass inflow in the order of $10^{-14}$ grams, which does not appreciably alter the bubble content \cite{Baghdassarian,Plesset,BrennerHilgenfeldtLohse}.
Therefore, contributions from (i) and (ii) are likely to be preponderant.
Moreover, these calculations show that a tenfold increase of air saturation in the water is synonym with approximately a ten times augmentation of the non-condensable gas partial pressure within the bubble.
Interestingly, these values do not match the bubble's internal pressure returned by the Keller-Miksis model \cite{KellerMiksis}, which predicts a 60\% increase of $p_\text{g0}$ from the least to the most saturated water exploited.
\par Although we cannot rule out the possibility that the differences in rebound size are simply a minimal damping effect of the non-condensable gas partial pressure within the bubble, we have reasons to consider an alternative scenario.
During the experiments, the vapour pressure of the water randomly varied by approximately 56 Pa, as a consequence of the water temperature variation of 0.3 degrees.
Although we could expect variations in the dissolved gas for the same reason, we assume them to be negligible in this $T_\infty$ range.
The effect of the non-condensable gas appears then to be dominant, as a relatively big augmentation of the vapour pressure with respect to the non-condensable gas partial pressure did not alter the observed trend in bubble dynamics.
These results align with predictions from the models proposed by \citeauthor{Fujikawa} \cite{Fujikawa} and \citeauthor{AkhatovLindauLauterborn} \cite{AkhatovLindauLauterborn}, who suggested that a small fraction of non-condensable gas in the bubble could hinder the vapour condensation during the collapse, playing a major role in the bubble dynamics.
Consequently, a small increment of the non-condensable gas may result in a larger rebound bubble compared to the one resulting from a relatively large increase of vapour.
\par Furthermore, while one might expect a decreasing luminescence signal energy with increasing gas content due to the gas cushioning effect on the collapse, and the possible entrainment of more vapor during collapse \cite{AkhatovLindauLauterborn}, leading to a larger partial vapor pressure that could quench the bubble collapse and decrease $E_\text{lum}$ \cite{BrennerHilgenfeldtLohse,MossZimmerman}, our results align with the observations of \citeauthor{Hiller} \cite{Hiller} and \citeauthor{Gompf} \cite{Gompf} on sonoluminescence.
They reported that light emission decreased with decreasing air saturation.
On the contrary, this is in contrast to the aforementioned works of \citeauthor{Baghdassarian} \cite{Baghdassarian} and \citeauthor{Wolfrum} \cite{Wolfrum}, who did not find any clear trend of luminescence with gas concentration.
However, it is important to note that they did not perform a statistical analysis of the collected data, as was done in this work.
One of the possible explanations worth considering is related to the growth of surface perturbations around the bubble \cite{BrennenFission,Plesset,SupponenLuminescence}.
These perturbations may diminish the light emission in single transient cavitation bubbles, similarly to what occurs in single bubble sonoluminescence \cite{BrennerHilgenfeldtLohse}.
A smaller non-condensable gas partial pressure, which may imply a smaller vapour partial pressure, allows for greater shrinkage of the bubble at collapse, promoting a larger deviation from sphericity in the final collapse stage \cite{Plesset}.
Consequently, this may lead to a less efficient compression of the bubble contents, particularly affecting the more sensitive luminescence phenomena.
\par In conclusion, the observed variation in the average dynamics of cavitation bubbles in water at typical air saturation levels is limited.
However, even minor alterations in the behavior of one single bubble could have a pronounced impact on applications susceptible to cavitation, such as cavitation erosion, which deserve more investigation.
Furthermore, these findings may help improving sonochemical applications, given the increased luminescence energy observed at bubble collapse in more saturated solutions.
Finally, these results could also be valuable for the development of more sophisticated numerical tools.

\begin{acknowledgments}
We acknowledge support by the MSCA-ITN of the EU Horizon 2020 Research and Innovation program (Grant Agreement No. 813766) and by the Swiss National Science Foundation (Grant No. 179018).
\end{acknowledgments}

\section*{Conflict of Interest}
The authors have no conflicts to disclose.

\section*{Author Contributions}
\textbf{Davide Bernardo Preso}: Conceptualization; Formal analysis; Investigation; Methodology; Validation; Visualization; Writing - original draft.
\textbf{Daniel Fuster}: Conceptualization; Formal analysis; Methodology; Supervision; Writing - review \& editing.
\textbf{Armand Baptiste Sieber}: Investigation; Methodology; Visualization; Writing - review \& editing.
\textbf{Mohamed Farhat}: Conceptualization; Formal analysis; Funding acquisition; Project administration; Supervision; Writing - review \& editing.

\section*{Data Availability Statement}
The data that support the findings of this study are available from the corresponding author upon reasonable request.

\nocite{*}
\bibliography{aipsamp}

\end{document}